\documentclass[prl,aps,twocolumn,showpacs,floatfix,superscriptaddress]{revtex4-1}
\usepackage{graphicx}
\usepackage{subfigure}
\usepackage{bm}
\usepackage{amsmath}
\usepackage{tabularx}
\usepackage{wasysym}
\usepackage{array}
\usepackage{multirow}
\usepackage[version=3]{mhchem} 
\usepackage{float}
\usepackage{color}
\usepackage{hyperref}
\usepackage{lipsum}
\usepackage{ulem}
\makeatletter
\newcommand{\colorcaption}[2][]{%
  \begingroup%
  \renewcommand{\@caption@fignum@sep}{ (color online). }%
  \caption[#1]{#2}%
  \endgroup%
}
\makeatother

\begin{document}
\title{Orientationally Misaligned Zipping of Lateral Graphene and  Boron Nitride Nanoribbons with Minimized Strain Energy and Enhanced Half-Metallicity}

\author{Jiang Zeng}
\affiliation{International Center for Quantum Design of Functional Materials (ICQD), Hefei National Laboratory for Physical Sciences at Microscale (HFNL), and Synergetic Innovation Center of Quantum Information and Quantum Physics, University of Science and Technology of China, Hefei, Anhui 230026, China}

\author{Wei Chen}
\affiliation{International Center for Quantum Design of Functional Materials (ICQD), Hefei National Laboratory for Physical Sciences at Microscale (HFNL), and Synergetic Innovation Center of Quantum Information and Quantum Physics, University of Science and Technology of China, Hefei, Anhui 230026, China}
\affiliation{Department of Physics and School of Engineering and Applied Sciences, Harvard University, Cambridge, Massachusetts 02138, USA}

\author{Ping Cui}
\affiliation{International Center for Quantum Design of Functional Materials (ICQD), Hefei National Laboratory for Physical Sciences at Microscale (HFNL), and Synergetic Innovation Center of Quantum Information and Quantum Physics, University of Science and Technology of China, Hefei, Anhui 230026, China}

\author{Dong-Bo Zhang}
\thanks{Corresponding authors:\\zhangzy@ustc.edu.cn\\dbzhang@csrc.ac.cn}
\affiliation{Beijing Computational Science Research Center, Beijing 100094, China}

\author{Zhenyu Zhang}
\thanks{Corresponding authors:\\zhangzy@ustc.edu.cn\\dbzhang@csrc.ac.cn}
\affiliation{International Center for Quantum Design of Functional Materials (ICQD), Hefei National Laboratory for Physical Sciences at Microscale (HFNL), and Synergetic Innovation Center of Quantum Information and Quantum Physics, University of Science and Technology of China, Hefei, Anhui 230026, China}


\begin{abstract}
Lateral heterostructures of  two-dimensional materials may exhibit various intriguing emergent properties. Yet when specified to the orientationally aligned heterojunctions of zigzag graphene and hexagonal boron nitride ($h$BN) nanoribbons, realizations of the high expectations on their properties encounter two standing hurtles. First, the rapid accumulation of strain energy prevents large-scale fabrication.  Secondly,  the pronounced half-metallicity predicted for freestanding graphene nanoribbons is severely suppressed. By properly tailoring orientational misalignment between zigzag graphene and chiral $h$BN nanoribbons, here we present a facile approach to overcome both obstacles. Our first-principles calculations show that the strain energy accumulation in such heterojunctions is significantly diminished for a range of misalignments. More strikingly, the half-metallicity is substantially enhanced from the orientationally aligned case, back to be comparable in magnitude with the freestanding case. The restored half-metallicity is largely attributed to the recovered superexchange interaction between the opposite heterojunction interfaces. The present findings may have important implications in eventual realization of graphene-based spintronics.
\end{abstract}


\maketitle

Recent research in the design, synthesis, and property characterization of two-dimensional (2D) heterostructures represents a major advance in low-dimensional materials science~\cite{vdw2}. As compelling examples, lateral heterostructures of graphene (G), hexagonal boron nitride ($h$BN), and other 2D materials~\cite{gbn1,gbn2,gbn3,gbn4,gbn5,gbn6,gbn7,gbn8,gbn9,gbn10,gbn11,mose1,mose2} have been successfully fabricated, where the 2D materials with distinctively different band structures are integrated by lateral covalent bonding within a single atomic layer. This new class of hybrid materials may exhibit novel and diverse properties~\cite{electron2,electron3,electron5,electron11}, which are expected to have a broad range of applications in electronic devices.

To date, extensive experimental~\cite{gbn4,gbn5,gbn6,gbn7,gbn8,gbn9,gbn10,gbn11} and theoretical~\cite{electron1,electron3} investigations have been devoted to orientationally aligned G-$h$BN heterojunctions. Whereas such heterostructures have clear advantages, there exist standing challenges in actual realization of the intriguing electronic and spintronic properties of graphene nanoribbons (GNRs). The first is related to the fabrication of the G-$h$BN heterostructures. It has been shown experimentally that small pieces of orientationally aligned G-$h$BN lateral heterojuctions can be formed with high degrees of crystallinity and abrupt interfaces~\cite{gbn4,gbn5,gbn6,gbn7,gbn8,gbn9}, but such coherently strained samples cannot be enlarged significantly, mainly due to their lattice mismatch. It is known that in epitaxial growth, the strain caused by lattice mismatch is nonlocal~\cite{epi1}, and the associated strain energy accumulates along with the domain size. As such,  the release of strain energy upon a critical domain size proceeds via initiation of structural instabilities such as dislocations~\cite{gbn10}. The other major and more fundamental challenge is associated with the preservation of the electronic and spintronic properties. Specifically, freestanding zigzag GNRs have been predicted to exhibit half-metallicity under a strong transverse electric field~\cite{hm,louie}. Because it has not been possible to apply the required strong external fields at such nanoscales, alternative approaches have been proposed to establish the required fields internally, for example via proper molecular adsorption~\cite{yang} or sandwiching a GNR between two $h$BN NRs~\cite{electron1,electron2,electron3,electron11}. However, even though the electric field associated with the charge transfer between the two inequivalent G-$h$BN boundaries is indeed strong~\cite{electron3}, the corresponding half-metallicity has been shown to be severely suppressed from that of freestanding GNRs~\cite{electron8}.

In this Letter, we provide a new scheme to overcome the main challenges described above, by invoking orientationally misaligned lateral heterojunctions consisting of zigzag G and chiral $h$BN NRs that otherwise satisfy structural commensurability conditions~\cite{cc1,cc2,cc3,cc4}. Using first-principles calculations, we show that the strain energy accumulation in such heterojunctions can be drastically diminished regardless of the domain size. More importantly, we reveal substantial enhancements in the half-metallicity from the orientationally aligned case, back to be comparable in magnitude with that of the freestanding GNR. Through detailed analysis of the spatial charge distribution, we attribute the restored half-metallicity to the enhancerecoveredd superexchange interaction, which greatly reinforces the coupling between the spin states at the two opposite interfaces.

\begin{figure}[tb]
\includegraphics[width=1\columnwidth]{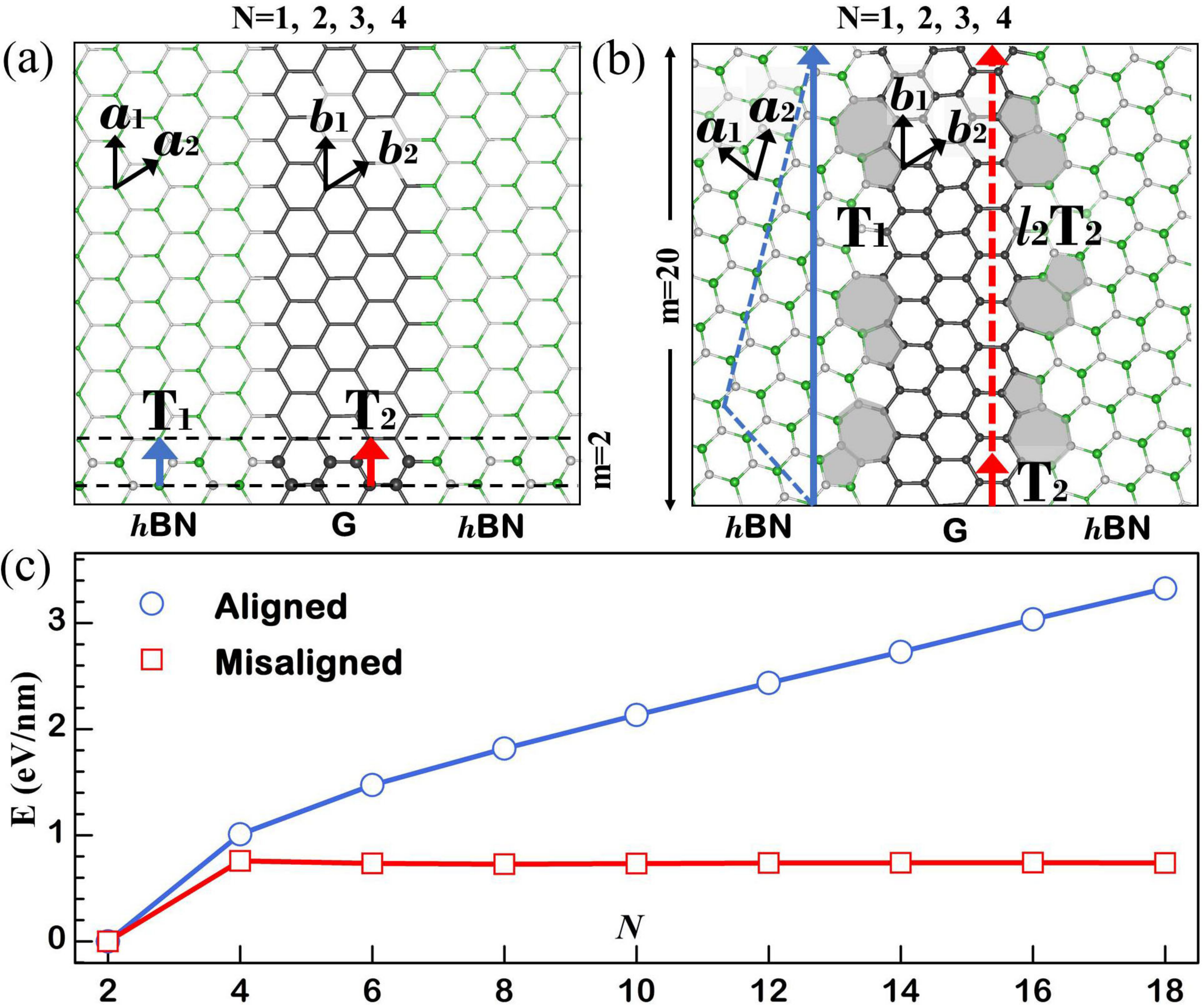}
\colorcaption{ Lateral heterojunctions of zigzag G and  $h$BN NRs with (a) orientational alignment and (b) orientational misalignment, where the GNRs contain $N=4$ zigzag C chains. The balls represent the atoms inside the computational supercell. For each C chain, there are $m=2$ and 20 C atoms in the unit cell for the (a) orientational aligned and (b) misaligned case, respectively. Lattice vectors ${\bf a}_1$ and ${\bf a}_2$ (${\bf b}_1$ and ${\bf b}_2$) define the hexagonal lattice for $h$BN (G). In (a) the dashed lines guide the unit cell. In (b) the dislocations formed at the interfaces are grey shaded. (c) Accumulations of the strain energy as a function of the GNR width measured by the number of zigzag C chains, $N$, for both orientationally aligned and  orientationally misaligned heterojunctions.}

\label{fig1}
\end{figure}

\begin{table}[tb]
\caption{Possible solutions to Eq.~(\ref{eq1}) under the constrain of  ${\bf T_2}=(1,0)$ for lateral heterojuctions consisting of zigzag GNR and chiral $h$BN NR.}
\centering
\begin{tabular}{p{0.08\linewidth}p{0.15\linewidth}<{\centering}p{0.26\linewidth}<{\centering}p{0.27\linewidth}<{\centering}p{0.12\linewidth}<{\centering}}
\hline
\hline
 $l_1$ & $l_2$  & $(n_1,n_2)$  & $\delta (\%)$ & $\theta (^\circ)$ \\
\hline
 $1$ & 10 & $(3,8)   $  & 0.26  &15.3\\
 $1$ & 11 & $(3,9)   $  & 0.10  &13.9\\
 $1$ & 12 & $(3,10) $  & 0.02  &12.7\\
$1$ & 13 &  $(3,11) $  & 0.02  &11.7\\
$1$ & 14 &  $(3,12) $  & 0.03  &10.9\\
$1$ & 15 &  $(3,13 )$  & 0.03  &10.2\\
\hline
\hline
\label{tb:Reference Vectors1}
\end{tabular}
\end{table}

The concept of commensurability matching has been widely invoked in studies of grain boundaries~\cite{cc1,cc3,cc4} and  heterostructures~\cite{cc2} in 2D materials. For  the zipping of zigzag G  and chiral $h$BN NRs, it simply reads,
\begin{equation}
\label{eq1}
l_1{\bf T_1}=l_2{\bf T_2},
\end{equation}
where, ${\bf T_1}$ and ${\bf T_2}$ are the primitive translation vectors along the interface for chiral $h$BN NR and zigzag GNR, respectively, as shown in Figs.~1(a) and 1(b). Here $l_1$ and $l_2$ are integer numbers, with $l_1=1$ for most cases.  Using the lattice vectors shown in Fig.~1(b), it is convenient to express  ${\bf T_1}=n_1{\bf a}_1+n_2{\bf a}_2$ or $(n_1,n_2)$ and ${\bf T_2}={\bf b}_1$ or $(1,0)$, with $|{\bf T_1}|=a_0\sqrt{{n_1}^{2}+n_1n_2+{n_2}^{2}}$ and $|{\bf T_2}|=b_0$, $n_1$ and $n_2$ are integer numbers, while $a_0=1.450$ {\AA} and $b_0=1.425$  {\AA} are the length of the lattice vectors for $h$BN and G, respectively. The orientational misalignment between $h$BN and G also equals the chiral angle of $h$BN NRs, which is defined as,
\begin{equation}
\label{eq2}
\theta=\tan^{-1}\big[\sqrt{3}n_2/(2n_1+n_2)\big].
\end{equation}
For the specific case shown in Fig.~1(b), ${\bf T}_1=(3,8)$, ${\bf T}_2=(1,0)$, and $\theta=15.3^{\circ}$. In general, Eq.~(\ref{eq1}) can only be approximately solved  such that a residue value in the lattice mismatch still remains,
\begin{equation}
\label{eq3}
\delta=\big|l_1|{\bf T_1}|-l_2|{\bf T_2}|\big|/(l_1|{\bf T_1}|).
\end{equation}
Table~\ref{tb:Reference Vectors1} lists the optimal solutions to Eq.~(\ref{eq1}) that minimize $\delta$ for a series of possible misalignment angles. We note that for the orientationally aligned case as shown in Fig.~1(a), the lattice mismatch is relatively large, given by $\delta=(a_0-b_0)/a_0\sim1.8\%$. In contrast, here it can be seen that the lattice mismatch is much smaller, with $\delta<0.3\%$. The  vanishingly small lattice mismatch indicates that the  commensurability conditions are well satisfied. We therefore can expect that the accumulation of strain energy in the orientationally aligned heterostructure can now be essentially avoided. This is of critical importance for the fabrication of large-scale G-$h$BN heterostructures that may exhibit various intriguing electronic/spintronic properties.

To quantitatively confirm these expectations on the energetic and electronic properties, we have carried out systematic first-principles density functional theory (DFT) studies of various G-$h$BN heterojunctions, including both orientationally aligned and misaligned cases. For the misalignment cases, we choose the heterostructures with the same misalignment angle of $\theta\approx15.3^{\circ}$ as shown in Fig.~1(b), but with varying GNR widths. We also note that this case corresponds to the largest residue mismatch, with all the other misalignments given in Table~\ref{tb:Reference Vectors1} exhibiting substantially lower residue mismatches. Our DFT calculations were made using the Vienna \textit{ab initio} simulation package  \begin{footnotesize}(VASP)\end{footnotesize}~\cite{dft1}, where the projector augmented plane wave (PAW) method~\cite{dft2,dft3}  was used, and the generalized gradient approximation (GGA) in the framework of Perdew-Burke-Ernzerhof (PBE)~\cite{dft4}  was adopted for the exchange-correlation interaction. The positions of the atoms were obtained by structural optimization until the forces on each atom are smaller than 0.01 eV/{\text {\AA}}. We have also taken the corrugations associated with out-of-plane atomic displacements into consideration, and  used a vacuum layer of 20 {\text {\AA}} to avoid the possible effects of image supercells. A plane-wave basis was set with a kinetic-energy cutoff of 500 eV, and the Brillouin zone was sampled by a $5\times5\times1$ or $5\times31\times1$ $k$-mesh, depending on the supercell size.

\begin{figure}[tb]
\includegraphics[width=1\columnwidth]{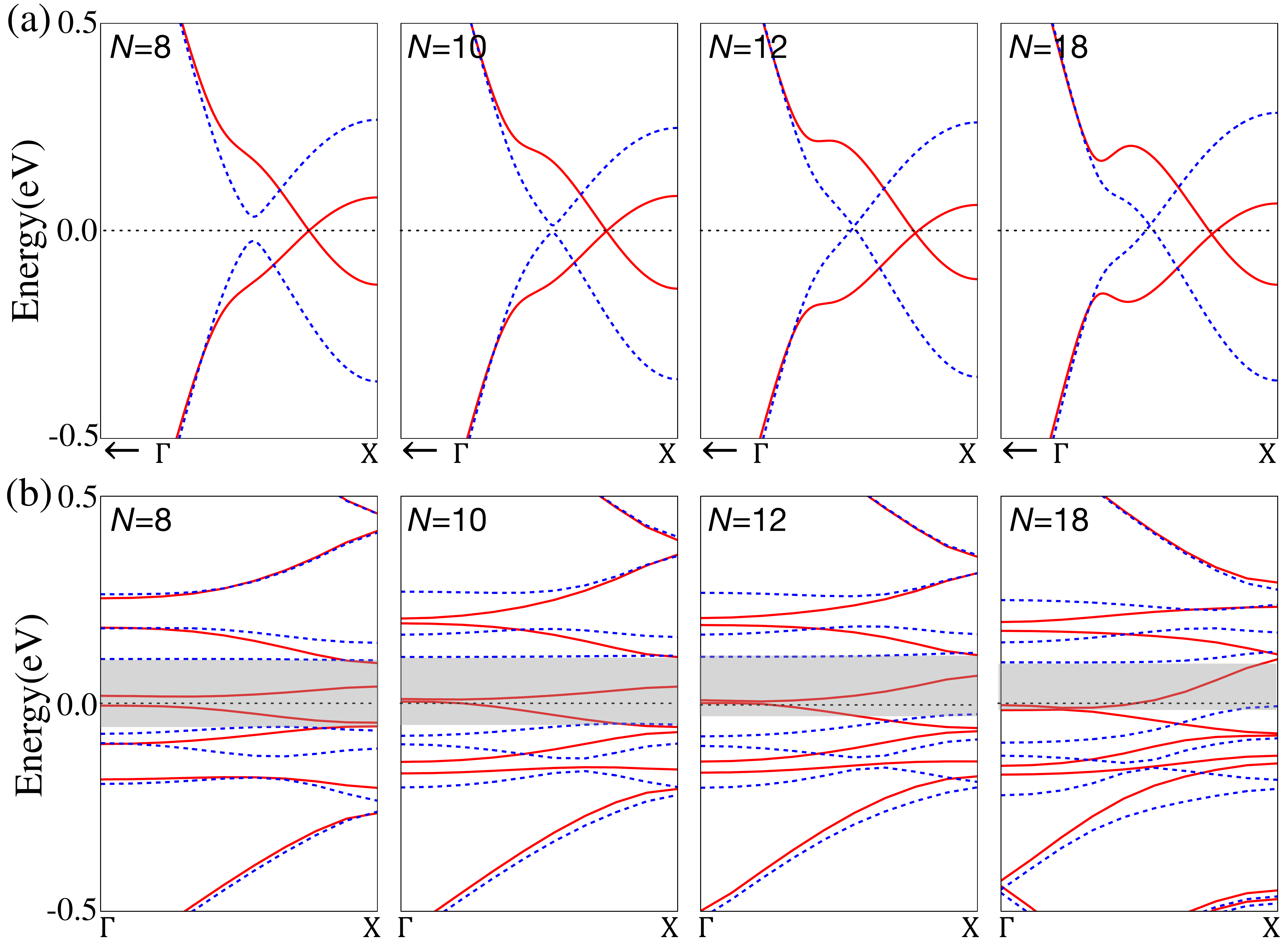}
\colorcaption{The spin-resolved band structures of orientationally (a) aligned and (b) misaligned lateral heterojunctions of zigzag G and $h$BN NRs for GNRs with different widths, $N$. The $\Gamma\rightarrow {\text X}$ direction corresponds to the interfacial direction spatially. The misalignment angle in (b) is $\theta\approx15.3^{\circ}$. The solid (blue) and dashed (red) curves represent different spin orientations. In (b) the half-metallic gap is grey shaded. The Fermi energy is set at zero as indicated by the horizontal  short dashed lines. }
\label{fig:band}
\end{figure}

For an effective characterization of the  strain energy accumulation in the GNR, the length of the supercell parallel to the interfacial direction is determined by the lattice constant of pristine $h$BN and fixed during the structural optimization, while the length perpendicular to the interfacial direction is optimized energetically. We calculate the strain energy versus the GNR width in terms of the number of zigzag C chains, $N$, as,
\begin{equation}
\label{eq4}
E=E_N-E_2-m(N-2)E_{G},
\end{equation}
where $E_N$ or $E_2$ represents the  formation energy of the heterostructure with $N$ or $2$ carbon chains, respectively, $E_{G}$ denotes the formation energy of pristine graphene per C atom, and $m$ is the number of C atoms in one C chain contained in the supercell. Note that $m=2$ and $20$ for the heterostructures shown in Figs.~1(a) and 1(b), respectively, while $E_N$, $E_2$, and $E_G$ are all obtained from detailed first-principles calculations.  In particular, the term ($E_N-E_2$) cancels out not only the contributions from the respective $h$BN components in orientationally aligned and misaligned cases, but also the contributions from the formation of the C-B and C-N bonds at both interfaces for either alignment. This choice ensures that the strain energy $E$ contains purely the energy raised from the enlargement or ``growth" of the GNR width. Finally, $E$ is normalized to energy per unit length along the interfacial direction.

The results are summarized in Fig.~1(c). First of all, for both the orientationally aligned and misaligned cases, the strain energy increases rapidly for $N\leq4$, largely due to the local structural relaxation near the interfaces. In the $N>4$ region, distinctly different behaviors of the strain energy are identified. For the orientationally aligned case, the strain energy increases monotonously with a near-linear dependence on $N$, indicating rapid  accumulation of the strain energy. This finding is consistent with the recent experimental revelation~\cite{gbn10} that  defects such as dislocations, discontinuities, and ripples emerge to release the accumulated strain energy in the heterostructure of G-$h$BN during the epitaxial growth.  On the other hand, for the orientationally misaligned case, it is reassuring to see that the strain energy indeed stays at a relatively low constant level, independent of $N$.

The preceding analysis reveals that the superior structural stability can be achieved in properly misaligned heterojunctions of zigzag G and chiral $h$BN NRs, where the accumulation of strain energy can be effectively diminished. As a consequence, heterojunctions at relatively large scales can be fabricated. The next question is how such intentional misalignments would influence the physical properties of the heterostructures, which contain periodic arrays of dislocations in the form of pentagon-heptagon pairs along the interfaces [see Fig.~1(b)]. In this regard, we note that similar structural topologies in polycrystalline graphene have been found to enhance the mechanical stability~\cite{vivek}, to alter the transport properties across the grain boundaries~\cite{cc1}, or to introduce emergent magnetic properties along the boundaries~\cite{castro neto}.

\begin{table}
\caption{Comparison of the half-metallic gaps (meV) of orientationally aligned heterojunctions,  misaligned heterojunctions, and freestanding GNRs at different GNR widths, $N$. The results in Ref.~\cite{electron3} and Ref.~\cite{louie} are shown as references.}
\centering
\begin{tabular}{p{0.04\linewidth}p{0.17\linewidth}<{\centering}p{0.10\linewidth}<{\centering}p{0.25\linewidth}<{\centering}p{0.08\linewidth}<{\centering}p{0.10\linewidth}<{\centering}p{0.12\linewidth}<{\centering}}
\hline
\hline
{$N$} &\multicolumn{2}{c}{Aligned }& {Misaligned}&\multicolumn{3}{c}{Freestanding GNR}\\
\hline
  &This work& Ref.~\cite{electron3}&{This work} &\multicolumn{2}{c}{This work} & Ref.~\cite{louie}\\
&PBE &PBE&PBE& PBE&LDA&LDA\\
\hline
 8   & 75 & 71& 160  & 691  & 405 & 443  \\
10  & 18 & 0  & 160  & 629  & 362 &  --    \\
12  & 0   & 0  & 138  & 555  & 327 &  --    \\
14  & 0   & --  & 131  & 471  & 266 &  --    \\
16  & 0   & --  & 112  & 425  & 276 & 290 \\
18  & 0   & --  & 106  & 386  & 271 &  --   \\
\hline
\hline
\label{tb:gap}
\end{tabular}
\end{table}

\begin{figure}[tb]
\includegraphics[width=1\columnwidth]{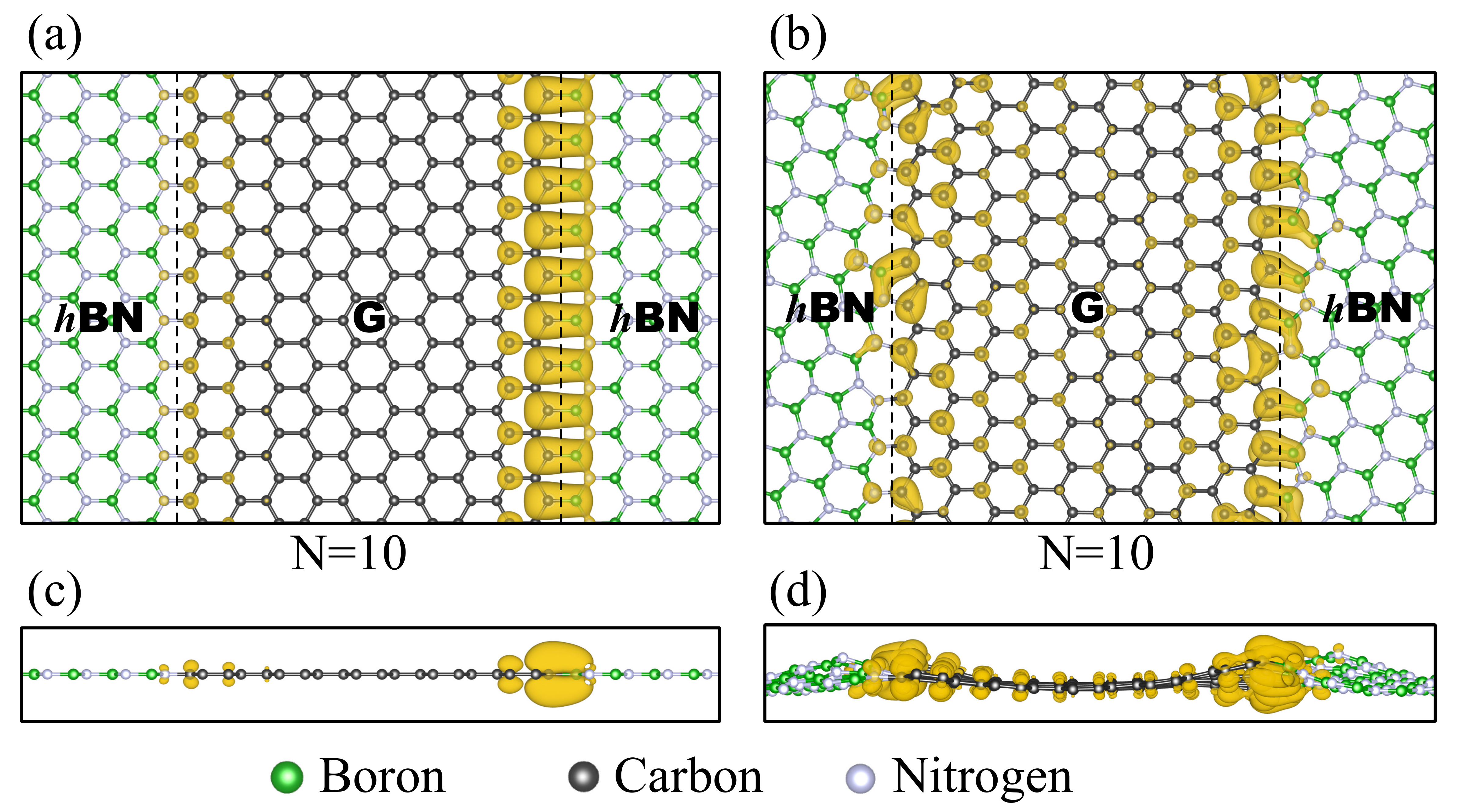}
\colorcaption{Spatial charge distributions of the edge electronic states within the energy range of $0.0-0.5$ eV below the Fermi level,  for both orientationally (a) aligned and (b) misaligned lateral heterojunctions of zigzag G and $h$BN NRs. The isosurface level is $0.01~{\text e/{\text \AA}}^3$. The misalignment angle in (b) is $\theta\approx 15.3^{\circ}$. (c) and (d) are the side view of the (a) and (b), respectively, displaying the vertical corrugations in the misaligned case. The black dashed lines guide the interfaces.}
\label{fig:electron}
\end{figure}

Figure~2 compares the band structures of orientationally aligned and misaligned G-$h$BN lateral heterojunctions for GNRs with different widths. Although the band structures are spin-polarized, and half-metallicity can be identified for both cases, the differences between the two are substantial.  For the aligned cases, the band structures display half-metallicity only for $N\leq10$. The half-metallic gap decays rapidly with $N$, and  is already nearly closed  when $N>10$, indicating an effective conventional metal, Fig.~2(a). On the other hand, for the misaligned cases, the band structures reveal pronounced half-metallicity for all the range of $N$ considered. The half-metallic gap is relatively large, and the decay with $N$ is not as significant, Fig.~2(b). We have also calculated quantitatively the half-metallic gaps of freestanding zigzag GNRs. The results are summarized in Table~\ref{tb:gap}. Compared to freestanding GNRs, the half-metallic gap is severely suppressed in the orientationally aligned heterojunctions, while restored greatly in the misaligned heterojunctions. These results can be rationalized by comparing with previous reports~\cite{electron3,louie}. For orientationally aligned heterojunctions and freestanding GNRs, our results show good agreement with those from the earlier studies. Of course, it should be noted that the PBE results are $\sim 40\%$ larger than the local density approximation (LDA) results, as the PBE functional overestimates the band gaps, while the LDA functional underestimates the band gaps.

The enhanced half-metallicity in the misaligned cases originates from the enhanced superexchange interaction between the opposite interfaces. This insight is obtained by analyzing the spatial charge distribution along the interfaces.  Recall that in freestanding zigzag GNR, the edge states are spin-polarized with ferromagnetic order along a given edge but are antiparallel for the two opposite edges~\cite{edge3,edge4,edge1,edge2}. These features are critical for achieving half-metallicity. In particular, due to the localization of the electronic density at either side of  a freestanding GNR~\cite{edge1,edge2}, the interedge superexchange interaction is pronounced~\cite{super,super2}. This superexchange interaction opens a gap for the ground states, and  at the presence of a strong external electric field, a sizeable  half-metallic gap can be generated. Unfortunately, such ideal spatial charge distributions cannot be readily preserved for the heterojunctioned systems. For the orientationally aligned cases, the charge density of the edge states at one interface is vanishingly small, Fig.~3(a), presumably due to the asymmetric bonding of C-B and C-N at the opposite interfaces~\cite{electron8}. Consequently, the superexchange interaction between the opposite interfaces is drastically diminished. In contrast, for the orientationally misaligned cases, the charge of the edge states is distributed inhomogeneously along the two interfaces due to the presence of the dislocations, Fig.~3(b). The charge densities in the misaligned cases are lower than that of freestanding GNRs~\cite{edge3,edge4,edge1,edge2}, but are comparable in magnitude along the two opposite interfaces, leading to the largely restored strengths in the superexchange interaction and half-metallicity.

\begin{table}[tb]
\caption{Comparison of $\triangle E$ (meV/nm) favoring AFM in the orientationally aligned heterojunctions, misaligned heterojunctions, and freestanding GNRs at different GNR widths, $N$.  }
\centering
\begin{tabular}{p{0.10\linewidth}<{\centering}p{0.20\linewidth}<{\centering}p{0.2\linewidth}<{\centering}p{0.35\linewidth}<{\centering}}
\hline
\hline
 $N$ & Aligned  & Misaligned  & Freestanding GNR \\
\hline
8   & 4 &  10 & 335  \\
10 & 2 &  8   & 341  \\
12 & 1 &  7   & 352  \\
14 & 1 &  6   & 334  \\
16 & 1 &  6   & 367  \\
18 & 1 &  6   & 372  \\
\hline
\hline
\label{tb:energy}
\end{tabular}
\end{table}

The enhanced superexchange interaction in the orientationally misaligned cases is further supported by the energy difference between the anti-ferromagnetic (AFM) and non-magnetic (NM) state,
\begin{equation}
\label{eq5}
\triangle E=E_{\text {NM}}-E_{\text {AFM}},
\end{equation}
where $E_{\text {AFM}}$ and $E_{\text {NM}}$ denote the energy of the AFM and NM states, respectively. The calculated $\triangle E$ is scaled to energy per nanometer in the interfacial direction.  Larger $\triangle E$ corresponds to stronger superexchange interaction. Thus, according to the data shown in Table~\ref{tb:energy}, the superexchange interaction is strongest for freestanding GNRs, and is stronger for the orientationally misaligned cases than the orientationally aligned cases.

We now briefly discuss the possibilities of experimentally realizing orientationally misaligned G-$h$BN lateral heterojunctions. We notice that when graphene is grown first, the $h$BN epilayer will take the crystallographic orientation of the G seed layer~\cite{gbn8}. On the other hand, it is feasible to obtain orientational misalignment between G and $h$BN by exploiting the different adhesive properties between G and $h$BN to a specific catalytic substrate such as Cu(100). It has been shown that the growth of $h$BN  undertakes  definitive crystallographic orientations following the square lattice of the Cu(100) surface~\cite{g1}, while the growth of G assumes rather random crystallographic orientations~\cite{gbn8,g2,g3}. Therefore, a specific orientational misalignment between G and $h$BN can be achieved by first selecting G with a proper crystallographic orientation from the randomly distributed samples grown on a Cu(100) substrate. Next, $h$BN can be grown at different areas on the Cu(100) substrate using mask approaches, and eventually coalesces with the G domain forming a misaligned G-$h$BN junction. Overall, given the clear advantages of such misaligned heterojunctions, we believe other more creative approaches will be devised on the experimental side for their realization.

In summary, we have introduced a new way to form zigzag G-$h$BN heterojunctions by properly tailoring the orientational misalignment between the two. The strain energy accumulation in such misaligned heterostructures is essentially eliminated, while the half-metallicity is found to be drastically enhanced from the orientationally aligned cases, back to be comparable in magnitude to that of freestanding GNRs. The pronounced half-metallicity is further attributed to the restored strength of the superexchange interaction between the electrons located at the two opposite interfaces. Overall, these revelations are valuable for potential practical realization of the intriguing emergent electronic and spintronic properties of G-$h$BN heterostructures.

This work was supported by the National Natural Science Foundation of China (Grant Nos. 61434002 and 11204286) and the National Key Basic Research Program of China (Grant No. 2014CB921103). The calculations were performed at the National Supercomputing Center in Shenzhen.

\end{document}